\documentclass[a4paper]{article}

\usepackage{INTERSPEECH2021}

\usepackage[group-separator={,}]{siunitx}

\usepackage{color}
\definecolor{todoColor}{rgb}{1,0,1}

\usepackage{hyperref,enumitem}

\def\mbf#1{\mathbf{#1}}

\def\bs#1{\boldsymbol{#1}}

\title{Learning robust speech representation \\ with an articulatory-regularized variational autoencoder}
\name{
Marc-Antoine Georges$^{1,2}$,
Laurent Girin$^1$,
Jean-Luc Schwartz$^1$,
Thomas Hueber$^1$
}
\address{
  $^1$Univ. Grenoble Alpes, CNRS, GIPSA-lab, 38000 Grenoble, France.\\
  $^2$Univ. Grenoble Alpes, CNRS, LPNC, 38000 Grenoble, France.}
\email{marc-antoine.georges@grenoble-inp.fr}

\begin{document}

\maketitle
\begin{abstract}

It is increasingly considered that human speech perception and production both rely on articulatory representations.
In this paper, we investigate whether this type of representation could improve the performances of a deep generative model (here a variational autoencoder) trained to encode and decode acoustic speech features.
First we develop an articulatory model able to associate articulatory parameters describing the jaw, tongue, lips and velum configurations with vocal tract shapes and spectral features. Then we incorporate these articulatory parameters into a variational autoencoder applied on spectral features by using a regularization technique that constraints part of the latent space to follow articulatory trajectories.
We show that this articulatory constraint improves model training by decreasing time to convergence and reconstruction loss at convergence, and yields better performance in a speech denoising task.

\end{abstract}

\noindent\textbf{Index Terms}: Speech production, representation learning, variational autoencoder, articulatory model, speech enhancement.

\section{Introduction}

Motor and perceptuo-motor theories of speech perception involve internal motor simulation processes \cite{Liberman1985, schwartz2012perception} which may be particularly recruited when perceiving speech in adverse (e.g. noisy) conditions \cite{SKIPPER201777}. Similarly, most computational models of speech motor control rely on an explicit access to motor representations, first when recovering the motor commands required to reach an acoustic target (inverse internal model), and then when simulating the acoustic consequences of articulatory gestures (direct internal model) \cite{Houde2011, Guenther2012}.
Inspired by human cognition and neurophysiology, the integration of motor/articulatory priors and constraints in automatic speech processing systems has motivated several studies, for increasing the  robustness of  automatic speech recognition (ASR) systems in noise \cite{Rose1996, Castellini2011, King2007}, for allowing a better control of  text-to-speech (TTS) synthesis \cite{ling_integrating_2009}, or for designing voice restoration and pronunciation training systems \cite{schultz_etal_taslp_2017_review_biosignal}.
Modeling the complex relationships between phonetic targets, articulatory movements and speech acoustics is also essential to build a computational model of speech perception and production  \cite{Laurent2017, HueberNC2019}.

\begin{figure}[t!]
  \centering
  \includegraphics[width=.8\linewidth]{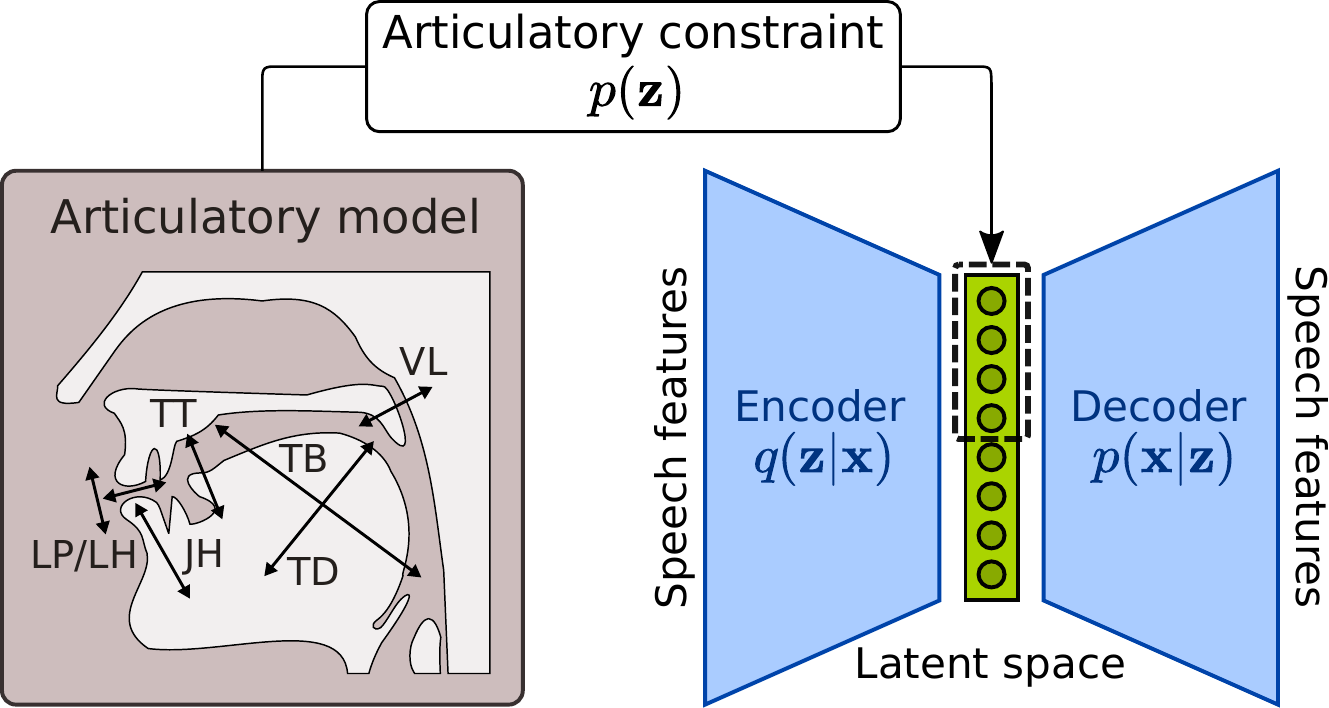}
  \caption{
    Schematic view of the proposed articulatory-regularized variational autoencoder.
  }
  \label{fig:overview}
\end{figure}

With both these technological and fundamental research goals in mind, we deal in this paper with the automatic learning of latent representation from the raw speech audio signal. We focus on deep generative models and in particular on the variational autoencoder (VAE) model \cite{Kingma2014,rezende2014stochastic} which can be seen as a probabilistic version of a deep autoencoder. The VAE has shown to be able to learn relevant latent representations by disentangling dimensions like speaker identity or phonetic features \cite{blaauw2016modeling, Hsu2017}. This model has already been successfully used in a variety of speech processing applications, e.g. \cite{hsu2016voice, bando2017statistical, akuzawa2018learning, leglaive2019semi}. In line with speech perception theory, we propose in this paper to introduce prior articulatory information in the representation learning process. This information is first derived from an articulatory model built from in-vivo recordings of a reference speaker. It is then transferred at training time via an additional regularization term in the loss function of the VAE. An overview of the proposed architecture is shown in Figure \ref{fig:overview}.
With this model, we address the two following research questions: 1) Can prior articulatory knowledge fasten the speech representation learning process? 2) Can prior articulatory knowledge make the learned latent representation more robust to noise? To address those questions, we compared the proposed articulatory-regularized VAE to a conventional one in term of convergence speed at training time and on a speech denoising task.

To the best of our knowledge, the introduction of prior articulatory constraints for representation learning has been proposed only in very few studies. In \cite{Turrisi2019}, a set of vocal tract variable derived from the articulatory phonology theory \cite{Ohala1986} is used to constrain the latent space of a (deterministic) auto-encoder. The performance is evaluated by measuring the accuracy of the reconstructed articulatory features whereas in the present study we focus on the quality of the reconstructed  speech signal. In \cite{Saha2020}, a normalizing flow technique is used to constraint the latent space of two autoencoders respectively processing articulatory and audio data. However, evaluation is mostly limited to qualitative evaluation. Therefore, the present study proposes the first VAE model regularized by articulatory prior knowledge and used to learn robust latent representation from the audio speech signal.

\section{Methodology}

\subsection{Acoustic and articulatory data}
\label{subsec:data}

The following experiments were conducted on two datasets. Each of these datasets is composed of parallel audio and EMA recordings (sustained vowels, vowel-consonant-vowel sequences, words, sentences).
The first dataset, PB2007, consists of 1,109 items (15 minutes of speech) produced by a reference speaker (PB, male).
EMA data were recorded using the Cartens 2D EMA system (AG200).
Six coils were placed on the tongue tip, blade and dorsum, upper and lower lip and on the jaw (lower incisor).
The acquired trajectories were low-pass filtered at 20~Hz and down-sampled from 200~Hz to 100~Hz. We denote by $\mathbf{y}$ an individual resulting vector of EMA data, of dimension 12.
The second corpus, BY2014,\footnote{available online at http://doi.org/10.5281/zenodo.154083} includes 925 items (45 minutes of speech) produced by another reference speaker (BY, male).
Articulatory trajectories were recorded using the 3D NDI Wave system with 9 coils used (3 on the tongue, 4 on the lips, 1 on the jaw, 1 on the velum). They were low-pass filtered at 20~Hz and down-sampled from 200~Hz to 100~Hz. The 3D coordinates of the 9 EMA coils were finally projected on the midsagittal plane resulting in a 14-dimensional vector $\mathbf{y}$.

As required by LPCNet vocoder \cite{valin2019lpcnet} which is used in this study to reconstruct an audio speech signal from the output of the VAE, each audio recording was converted into a sequence of 18 Bark-scale cepstral coefficients \cite{Schroeder1979}, using a 20-ms sliding analysis window with 10-ms frame shift. The resulting vector of audio features is denoted $\mathbf{x}$. 

\subsection{Building the articulatory model}
\label{subsec:articulatory}

For each of the two reference speakers, we first built an articulatory model using the general methodology originally proposed by \cite{Maeda1990}, slightly adapted in \cite{Serrurier2012} in order to process EMA articulatory data.
In the present study, we consider the six following  articulatory parameters (represented in Figure \ref{fig:overview}):
\textit{Jaw Height} (\textit{JH}),
\textit{Tongue Body} (\textit{TB}),
\textit{Tongue Dorsum} (\textit{TD}),
\textit{Tongue Tip} (\textit{TT}),
\textit{Lip Protusion} (\textit{LP}),
\textit{Lip Height} (\textit{LH}), \textit{Velum} (\textit{VL}, for BY speaker only). Here ``articulatory parameters'' means that those parameters are interpretable in terms of articulatory control/function in speech production \cite{Maeda1994}.
The core principle is to extract the latent dimension of tongue and lips movements after removing the contribution of the jaw, using a so-called ``guided-PCA''.
More precisely, the \textit{JH} parameter (and the  corresponding value for all the articulatory observations of the dataset) is defined as the first principal component (PC) of the jaw movement.
The contribution of the jaw to the movement of the tongue is then estimated using a linear regression between \textit{JH} values and the coordinated of the 3 EMA coils attached to the tongue. Once this contribution estimated and removed, the parameters \textit{TB} and \textit{TD} are defined as the two first PCs of the joint movement of tongue dorsum and back.
A linear regression between the  \textit{TB} and \textit{TD} on one hand and the tongue tip coils on the other hand provides a residual movement of the tongue, freed from the contribution of the jaw, the tongue dorsum and the tongue back. The \textit{TT}  parameters is finally defined as the first PC of this residual movement. A similar procedure is used for extracting the \textit{LP} and \textit{LH} lip parameters and corresponding values for both datasets. For the BY2014 corpus, the \textit{VL} parameter is simply defined as the first PC of the EMA coils attached to the velum.
In summary, at the end of this guided PCA analysis, we have a linear transformation to go from a vector of EMA parameters $\mathbf{y}$ to an articulatory vector $\mathbf{a} = [\textit{JH}, \textit{TB}, \textit{TD}, \textit{TT}, \textit{LP}, \textit{LH}]$ (for the PB2007 dataset) or $\mathbf{a} = [\textit{JH}, \textit{TB}, \textit{TD}, \textit{TT}, \textit{LP}, \textit{LH}, \textit{VL}]$ for the BY2014 dataset, and vice versa.
For convenience in the following, we denote by $\mathbf{a}(\mathbf{x})$ the articulatory vector corresponding to the cepstral vector $\mathbf{x}$. Because this set of parameter is reduced in size compared to EMA parameters, the articulatory information is ``compressed'' into a low-dimensional vector, which is appropriate for our later application of an articulatory constraint on the VAE latent space (which is also expected to be of low-dimension).

\subsection{The VAE model}

The seminal VAE model introduced in \cite{Kingma2014,rezende2014stochastic} is defined by:
\begin{equation}
p_{\bs{\theta}}(\mbf{x}, \mbf{z}) = p_{\bs{\theta}}(\mbf{x} | \mbf{z})p(\mbf{z}),
\label{joint_dist_y_z}
\end{equation}
where $p(\mbf{z})$, the prior distribution of the latent vector $\mbf{z}$, is a multivariate standard Gaussian distribution, $p_{\bs{\theta}}(\mbf{x} | \mbf{z})$ is the (conditional) likelihood function of the observed variable $\mbf{x}$, and the dimension $L$ of $\mbf{z}$ is (possibly much) lower than the dimension $F$ of $\mbf{x}$. The parameters of $p_{\bs{\theta}}(\mbf{x} | \mbf{z})$ are provided by a deep neural network (DNN), called the decoder network, that takes $\mbf{z}$ as input. $\bs{\theta}$ represents the parameters of this decoder network (e.g., the weights and biaises of a multi-layer perceptron). In the present work, $p_{\bs{\theta}}(\mbf{x} | \mbf{z})$ is a Gaussian distribution with diagonal covariance matrix.

Because the relationship between $\mbf{z}$ and $\mbf{x}$ is highly non-linear, the posterior distribution $p_{\bs{\theta}}(\mbf{z} | \mbf{x})$ is not analytically tractable. It is thus approximated with a parametric variational distribution $q_{\bs{\phi}}(\mbf{z} | \mbf{x})$, a.k.a.\ the inference model, whose parameters are provided by another DNN (called the encoder network, with weights $\bs{\phi}$ and input $\mbf{x}$).
A usual choice, that we follow here, is to set $q_{\bs{\phi}}(\mbf{z} | \mbf{x})$ as a Gaussian distribution with diagonal covariance matrix.
The parameters $\{\bs{\theta},\bs{\phi}\}$ are then jointly estimated by  maximizing a lower bound of the data log-likelihood function, called the Variational Lower Bound (VLB), given by (for one single data vector):
\begin{align}
\mathcal{L}(\bs{\phi}, \bs{\theta}, \mbf{x}) &=  \mathbb{E}_{q_{\bs{\phi}}(\mbf{z} | \mbf{x})}\big[ \log p_{\bs{\theta}}(\mbf{x} | \mbf{z}) \big] \! - \! D_{\text{KL}}\big[q_{\bs{\phi}}(\mbf{z} | \mbf{
x}) \parallel p(\mbf{z})\big], \label{eq:VAE-VLB-a}
\end{align}
and evaluated on a training dataset ($D_{\text{KL}}$ denotes the Kullback-Leibler divergence). The left term of the VLB represents the reconstruction accuracy of the encoding-decoding process and the right term is a regularization term that ensures some degree of ``disentanglement'' of the latent vector entries \cite{Kingma2014}. Maximization of the VLB is done by combining stochastic gradient descent with sampling techniques.

\subsection{The articulatory-regularised VAE (AR-VAE)}

In the present work, each vector $\mbf{x}$ is a vector of Bark-scale cepstral coefficients, of dimension 18, extracted from the audio as described in Section~\ref{subsec:data}. The dimension of $\mathbf{z}$ will be specified later.
To force the latent space of the VAE to fit the articulatory space, we added a third term to the above VLB using the same regularization technique as in \cite{roche2021} (itself inspired from \cite{Esling2018}):
\begin{align}
\mathcal{L}(\phi, \theta, \mathbf{x})
= &~ \mathbb{E}_{q_\phi(\mathbf{z} | \mathbf{x})}\big[\log\,{p_{\theta}(\mathbf{x} | \mathbf{z})}\big] - D_{\text{KL}}\big[q_\phi(\mathbf{z} | \mathbf{x}) \parallel p(\mathbf{z})\big] \nonumber
\\
& \qquad + \alpha \, \mathbb{E}_{q_\phi(\mathbf{z} | \mathbf{x})}\big[\mathcal{R}(\mathbf{z}, \mathbf{a}(\mathbf{x}))\big]. \label{eq:VLB-regularized}
\end{align}
The new regularization term  $\mathcal{R}(\mathbf{z}, \mathbf{a}(\mathbf{x}))$ ensures that for each speech frame the first entries of the latent vector $\mathbf{z}$ remain close to the corresponding entries in the vector of articulatory parameters  $\mathbf{a}(\mathbf{x})$ defined in Section~\ref{subsec:articulatory}.
In our experiments, we implemented this regularization term with the mean squared error (MSE):
\begin{equation}
	\mathcal{R}\big(\mathbf{z}, \mathbf{a}(\mathbf{x})\big) = \ \parallel \mathbf{z}_{1:N} - \mathbf{a}(\mathbf{x}) \parallel ^2, \label{eq:metric}
\end{equation}
where $\mathbf{z}_{1:N}$ denotes the subvector made of the $N$ first latent values, with $N=6$ or $7$ depending on the dataset. This term can be interpreted in statistical modeling terms as an additional Gaussian prior on $\mathbf{z}_{1:N}$ with mean vector $\mathbf{a}(\mathbf{x})$ and an arbitrary fixed variance.
In practice, the two expectations in \eqref{eq:VLB-regularized} are replaced with estimates based on Monte Carlo sampling of $\mathbf{z}$ (using the well-known reparameterization trick, just as in the seminal VAE \cite{Kingma2014}).
Finally, $\alpha$ is a weighting factor monitoring the weight of the articulatory regularization term, and that will be varied in our experiments. 

\subsection{Implementation}

We implemented the proposed articulatory-regularized VAE model with the following architecture: 
The encoder was composed of 4 fully connected hidden layers (256, 128, 64 and 32 neurons) and the decoder had the same, but reversed, composition. The size of the latent space was 12 (i.e. the size of the output layer of the encoder) when the model was trained on the PB2007 corpus (including 6 articulatory-regularized dimensions), and it was 14 when the model was trained on the BY2014 corpus (including 7 articulatory-regularized dimensions).
Note that half of the $\mathbf{z}$ entries are articulatory-constrained, hence they are forced to encode the information in cepstral vectors that is strongly correlated to the articulatory parameters, and the other half are let free to encode ``everything else'' (e.g. speech source information).

The hyperbolic tangent activation function was used for each hidden layer.
Model training was done using back-propagation with Adam optimizer, on mini-batches of 32 observations (pairs of $\mathbf{x}$ and $\mathbf{a}(\mathbf{x})$ vectors).
For each experience, the datasets were randomly partitioned with 80\% of the data used for training and the remaining 20\% used for testing.
The implementation was done using the  \textit{PyTorch} toolkit \cite{pytorch}.

\section{Experiments}

\subsection{Model learning speed and accuracy}

\begin{figure}[t!]
  \centering
  \includegraphics[width=\linewidth]{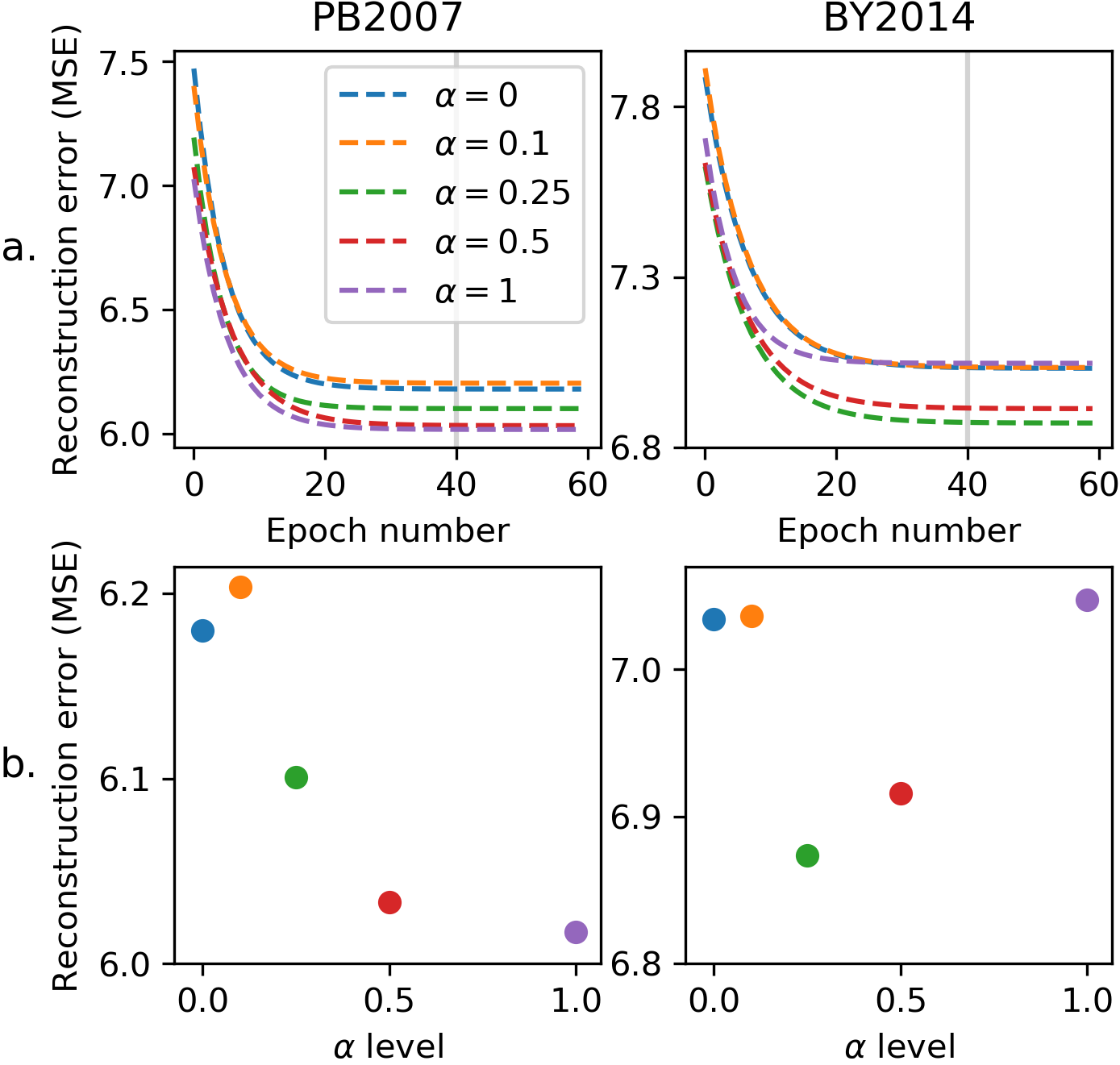}
  \caption{
    a.: Evolution of the reconstruction loss on the test set during training.
    For better visualization, each learning curve is fitted by an exponentially decreasing function
    b.: Final performance after convergence at epoch 40 on the test set.
  }
  \label{fig:training_alpha}
\end{figure}

We first tested if the introduction of articulatory constraints  could fasten the training process. To that purpose, for each of the two datasets, and for each value of $\alpha$ taken in $\{0, 0.1, 0.25, 0.5, 1\}$  we trained 10 AR-VAE models (with a different initialization each time) during 60 epochs (note that an AR-VAE with $\alpha=0$ is equivalent to a conventional VAE). At each epoch, we computed the reconstruction error defined as the MSE between the reconstructed and true cepstral coefficients, on the test set. For each value of $\alpha$ (and for each dataset), we finally built a smooth version of the learning curve by averaging the reconstruction loss on the test set, over the 10 models.
These learning curves are presented in Figure \ref{fig:training_alpha}a. First,  we observe that almost all AR-VAEs converge faster than the conventional VAEs (i.e. the blue dashed line is almost always above the other lines). Then, as summarized in Figure \ref{fig:training_alpha}b, the best final performance is obtained with the proposed AR-VAE on both datasets (i.e. with  $\alpha = 1$ for PB2007 and with $\alpha = 0.25$ for BY2014).  
Therefore, adding  articulatory constraints  does improve representation learning, both in terms of convergence speed and  accuracy. 

\subsection{Robustness to noise}

\begin{figure}[t!]
  \centering
  \includegraphics[width=0.9\linewidth]{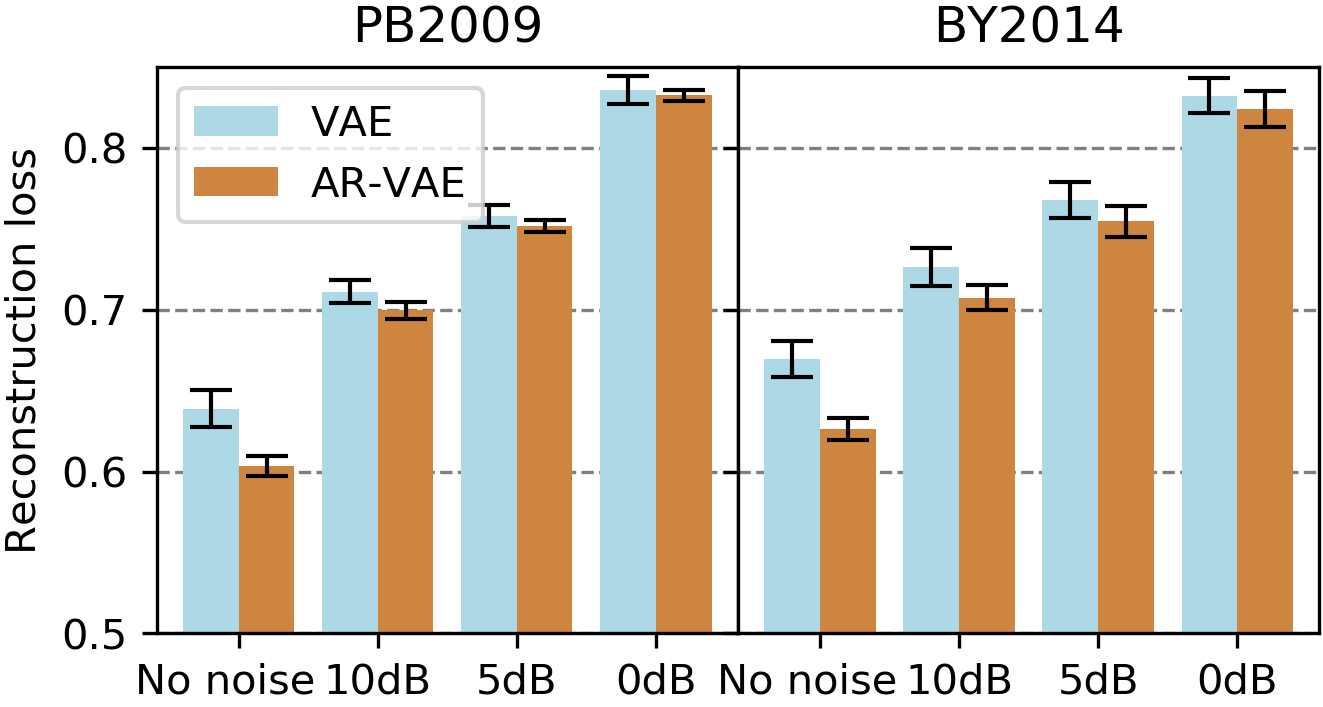}
  \caption{
    Reconstruction loss (mean $\pm$ standard deviation) of the conventional VAE and the proposed AR-VAE on the test set for the speech denoising task.
  }
  \label{fig:denoising_loss}
\end{figure}

We then tested the performance of the proposed AR-VAE on a speech denoising task.
To that purpose, a babble noise was added to each audio speech signal with different signal-to-noise ratios (SNR) (no noise, 10dB, 5dB and 0dB).
Sequence of acoustic feature vectors (i.e. 18 Bark-scale  cepstral coefficients) were extracted from the noisy audio signals with the same method used previously. For each dataset, both VAE and AR-VAE were trained to reconstruct the non-noisy version of each acoustic feature vector from its noisy counterparts.
For the AR-VAE, we first compared different values of the $\alpha$ parameter for this denoising task. For concision, we report here only the results with $\alpha=1$ which provided the best performance among the tested values.
As for the first experiment, we trained 10 different VAEs/AR-VAEs (with a different initialization each time) and averaged the results over the 10 runs. The reconstruction error on the test dataset is shown in Figure \ref{fig:denoising_loss}.
These results show that the proposed AR-VAE outperforms the conventional VAE on the denoising task, for all considered SNR. The performance difference is more pronounced for lower noise levels.

\begin{figure}[t!]
  \centering
  \includegraphics[width=0.9\linewidth]{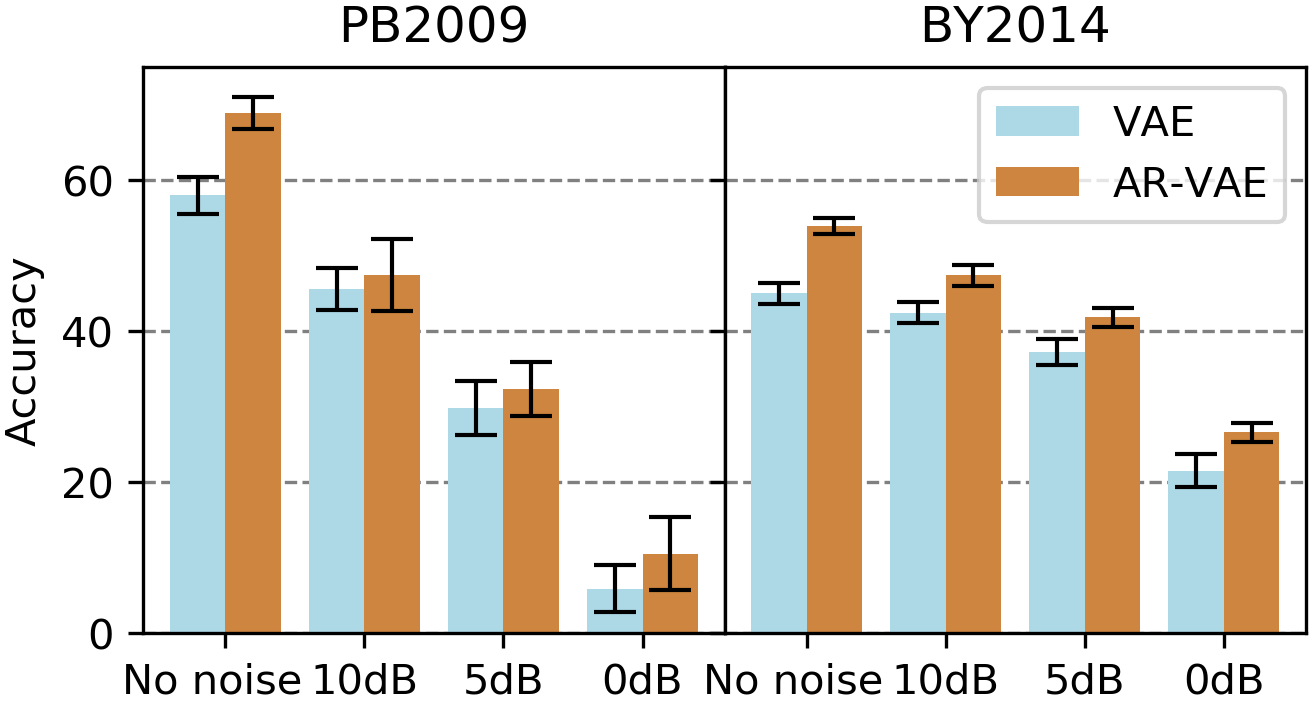}
  \caption{
    Accuracy (mean $\pm$ standard deviation) of an HMM-based phonetic decoder when processing  speech signals denoised by the conventional VAE and the proposed AR-VAE (with $\alpha = 1$).
  }
  \label{fig:reco_hmm}
\end{figure}

Since the absolute value of the reconstruction loss (which is in an arbitrary unit) is difficult to interpret, we then conducted further evaluations to assess the quality of the denoised speech. First, we evaluate its phonetic content. To that purpose, for each sentence of the test set, and for each considered SNR, we re-synthesized a speech signal using the LPCNet neural vocoder fed by the cepstral coefficients reconstructed by the VAE/AR-VAE, together with the original pitch parameters (period and correlation, extracted from the clean speech sound).\footnote{Several sound examples are available at \url{https://georges.ma/p/ar-vae}}
The resulting speech signals were sent to a Hidden Markov Model (HMM) based phonetic decoder, trained on the original (clean) speech signals of the training set (left-to-right 3 emitting states context-independant HMM-GMM, trained using the HTK toolkit and a standard procedure, no language model used).
The decoding accuracy (which takes into account insertion and deletion errors) are presented in Figure \ref{fig:reco_hmm}. Again, the AR-VAE outperforms the VAE, by a large margin (up to 10\%) when processing clean speech, and by a smaller one when processing noisy speech.

\begin{figure}[t!]
  \centering
  \includegraphics[width=0.9\linewidth]{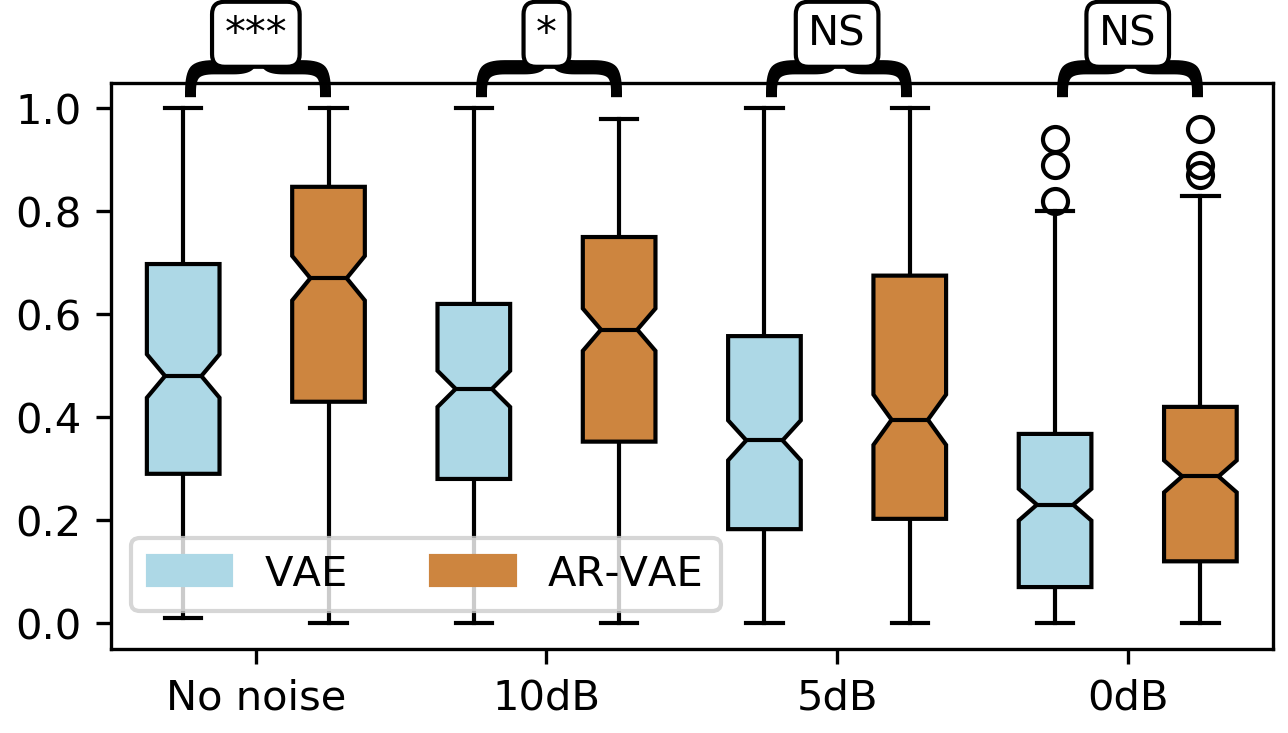}
  \caption{
    MUSHRA scores obtained for each level of noise and for the conventional VAE ($\alpha = 0$) and the proposed AR-VAE (with $\alpha = 1$).
    For the sake of clarity, we omit the anchor and reference scores.
    *** and * are significant ($p < 0.001$ and $p < 0.05$), and NS are non-significant differences.
  }
  \label{fig:mushra_results}
\end{figure}

Finally, we assessed the perceptive quality of the reconstructed speech signal using a MUSHRA test during which participants had to rank a set of audio stimuli by their similarity with a reference sound on a scale from 0 to 100~\cite{ITU2015}. 
We first randomly selected 20 short sentences from the BY2014 corpus (preferred to the PB2007 corpus because of the presence of data on the velum). For each sentence, we generated 6 audio stimuli: a low anchor built by adding babble noise to the original audio speech signal with a SNR of 0dB and re-synthesized (i.e.~vocoded) with LPCNet, a hidden reference built by re-synthesizing the original signal with LPCNet, a third stimulus built by first encoding-decoding the original signal either with the conventional VAE or with the proposed AR-VAE, and then synthesizing an audio signal with LPCNet, and four other stimuli built following the same principle but after having first added noise to the original signal with three levels of SNR ($10$dB, $5$dB and $0$dB).
We recruited 23 native French speakers online via the Prolific Academic platform~\cite{Palan2018}. Results are reported in Figure \ref{fig:mushra_results}. To assess the statistical significance of the difference between the MUSHRA scores, we first conducted a Kruskal-Wallis rank-sum test which showed a significant effect of the SNR factor ($p < 0.05$). Then a post-hoc Dunn test validated a statistically significant increase of performance from VAE to AR-VAE for clean audio ($p < 0.001$) and for noisy audio with SNR = 10dB, and showed no significant difference between the two models for SNR = 5dB and SNR = 0dB (i.e.~very noisy inputs).

\section{Discussion and perspectives}

The two experiments reported in Section 3 suggest that articulatory constraints improve the learning of speech representations in a VAE. This study opens interesting perspectives for assessing the combined role of articulatory knowledge and auditory processes in the elaboration of internal representations of speech signals.
While the efficiency of such articulatory-constrained VAE appears clearly in terms of spectrum reconstruction, it remains to explore how the corresponding representations are structured and whether they are able to integrate the concept of articulatory/motor invariance \cite{Liberman1967} inside phonetic representations.
We will particularly focus on the way articulatory information in the present VAE could improve the representation of plosive place of articulation, known to depend on the availability of articulatory/motor information  \cite{Barnaud2018}.

Another question concerns the way articulatory regularization provided by the articulatory data available for a given speaker could enable this speaker to better process the speech utterances of other speakers.
For this aim, we will explore how the regularized VAE performs in a denoising experiment involving multiple speakers. We will also study other VAE architectures in which articulatory-acoustic data for the speaking agent and acoustic-only data for other agents could be learnt together in articulatory-constrained VAE variants.

Finally, the precise developmental schedule along which a child develops her production and perception skills and learns the sounds of her language and the corresponding articulatory trajectories will shed interesting light on some algorithmic choices that should be made in further developments of such articulatory/motor constrained acoustic VAEs.

\section{Conclusion}

In this paper, we show how articulatory knowledge can help construct internal representations of auditory speech stimuli, by applying an articulatory regularization to VAEs encoding speech features, enforcing them to adopt mixed articulatory-acoustic representations in their latent space. The additional term in the training loss function enforcing a part of the latent spaces to follow articulatory parameters appears to improve learning efficiency, both in terms of learning speed and accuracy, and  also improve reconstruction performance in a denoising VAE. A number of perspectives are provided by this new type of joint articulatory-acoustic learning process.

\section{Acknowledgement}

This work has been partially supported by MIAI @ Grenoble Alpes (ANR-19-P3IA-0003).
The authors would like to thank Pierre Badin and Julien Diard for fruitful discussions.

\bibliographystyle{IEEEtran}

\bibliography{mybib}

\end{document}